\documentclass{webofc}
\usepackage[varg]{txfonts}   % Web of Conferences font
\usepackage{xcolor}
\pdfoutput=1

\begin{document}
\title{Maximum latent heat of neutron star matter without GR}
\author{\firstname{Eva} \lastname{Lope-Oter}\inst{1}
%\fnsep\thanks{\email {mariaevl@ucm.es}}
}
\institute{Dept. F\'{\i}sica Te\'orica and IPARCOS, Univ. Complutense de Madrid}

\abstract{%
We show how the specific latent heat is relevant to characterize the first-order phase transitions in neutron stars.
Our current knowledge of this dynamical quantity strongly depends on the uncertainty bands of Chiral Perturbation Theory and of pQCD calculations and can be used to diagnose progress on the equation of state. We state what is known to be hadron-model independent and without feedback from neutron star observations and, therefore,  they can be used to test General Relativity as well as theories beyond GR, such as modified gravity. 
}
\maketitle
%%%%%%%%%%%%%%%%%%%%%%%%%%%%%%%%%%%%%%%
\section{Introduction}
%%%%%%%%%%%%%%%%%%%%%%%%%%%%%%%%%%%%%%%

 One essential point in the present discussion on strongly interacting matter at finite density is whether there may be a first order phase transition.
 At the present time, we  do not know whether neutron-star matter undergoes such first-order phase transition to an exotic, perhaps non-hadronic phase, of great interest to nuclear and particle physics~\cite{Chesler:2019osn}. Many have been proposed, such as color-superconducting phases~\cite{Alford:2007xm}, inhomogeneous (crystalline-like) phases~\cite{Fulde:1964zz} or flavored~\cite{Oertel:2016xsn} or mixed phases~\cite{Heiselberg:1994fy} among others; but which or any presents itself in neutron stars remains under investigation~\cite{Llanes-Estrada:2019wmz}. 
 
 Observables including mass, radius, moment of inertia and tidal deformability  are largely determined by the equation of state that abstracts microscopic properties of the phase of the material. However, if there is a first-order transition between phases of very different energy density $\varepsilon$ then this could  verifiably affect the mass-radius relation, for example. Possible first-order phase transitions would leave distinct observable traces, such as a kink in the mass-radius diagram (accessible when neutron star radii become more routinely measured.
 
 A way to characterise the intensity of a phase transition is the latent heat that we here address.

%%%%%%%%%%%%%%%%%%%%%%%%%%%%%%%%%%%%%%%%%%%%%%%%%%
\section{Equations of state for Neutron Stars}
%%%%%%%%%%%%%%%%%%%%%%%%%%%%%%%%%%%%%%%%%%%%%%%%%%

Our approach consists in generating EoS ~\cite{Oter:2019kig} to be used to test GR as well as beyond GR, such as modified gravity. General Relativity  requires further testing because Einstein's equations $G^{\mu\nu} = -8\pi G T^{\mu\nu}$ have not been exhaustively constrained at the large $\varepsilon$ of neutron stars;  prior knowledge of the EoS $P( \varepsilon)$ from nuclear and hadron physics is needed,
 to have $T^{\mu\nu}$ inside NS purely from hadron theory without using astrophysics observables nor GR.
 Our EoS sets~\footnote{\tt teorica.fis.ucm.es/nEoS} are constrained only by input from hadron physics and fundamental principles, without feedback from neutron star observations. They are obtained starting from Chiral Perturbative theories (ChPT) at low density and ending at perturbative Quantum Chromodynamics (pQCD), using first principles (thermodynamic stability and causality) alone. These EoS sets are not as constrained as others in the recent literature~\cite{Godzieba:2020tjn}, but more reliable for testing  gravity. 
 
\begin{figure}[h]
\centering
\includegraphics[width=0.45\columnwidth]{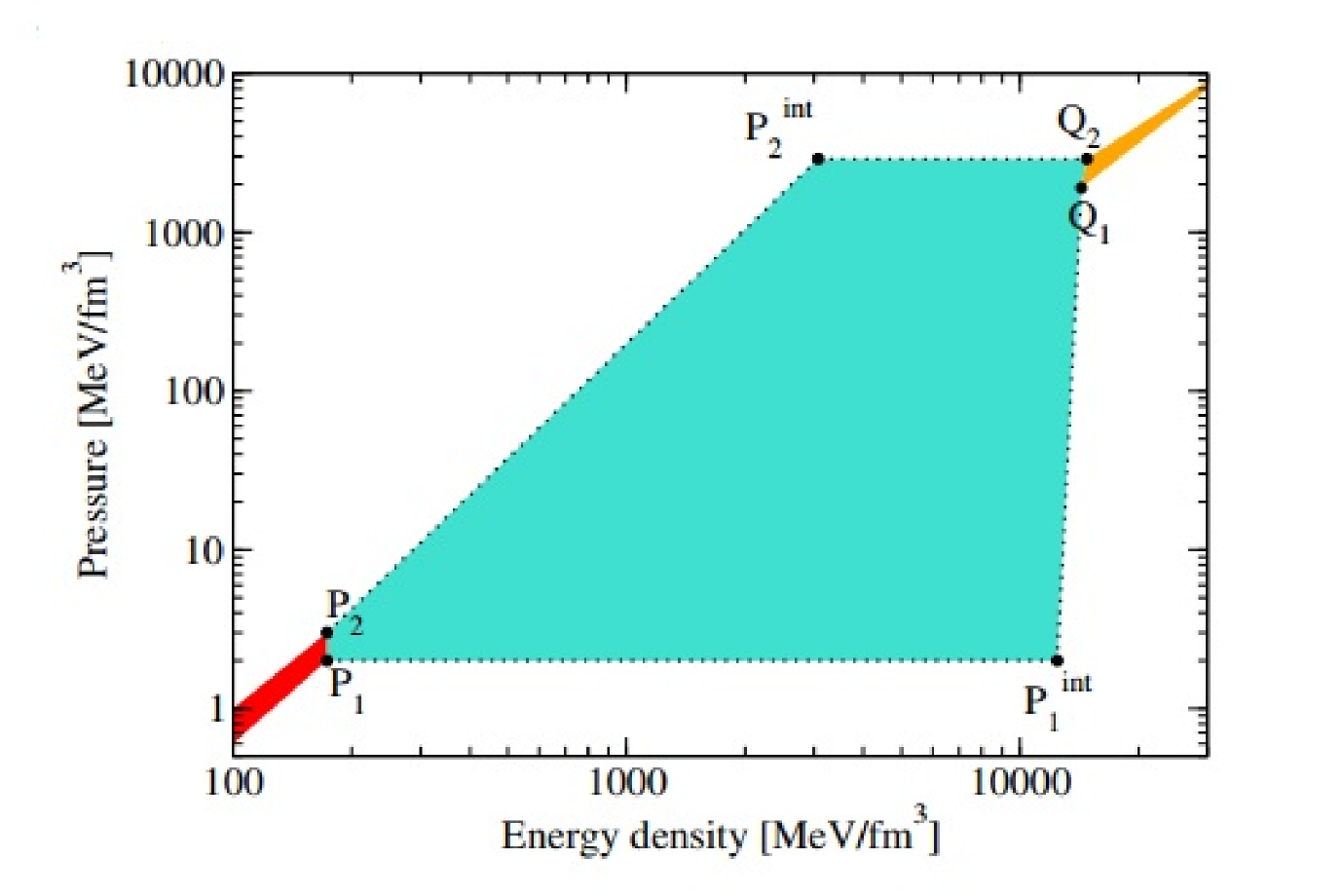}\ \ 
\includegraphics[width=0.45\columnwidth]{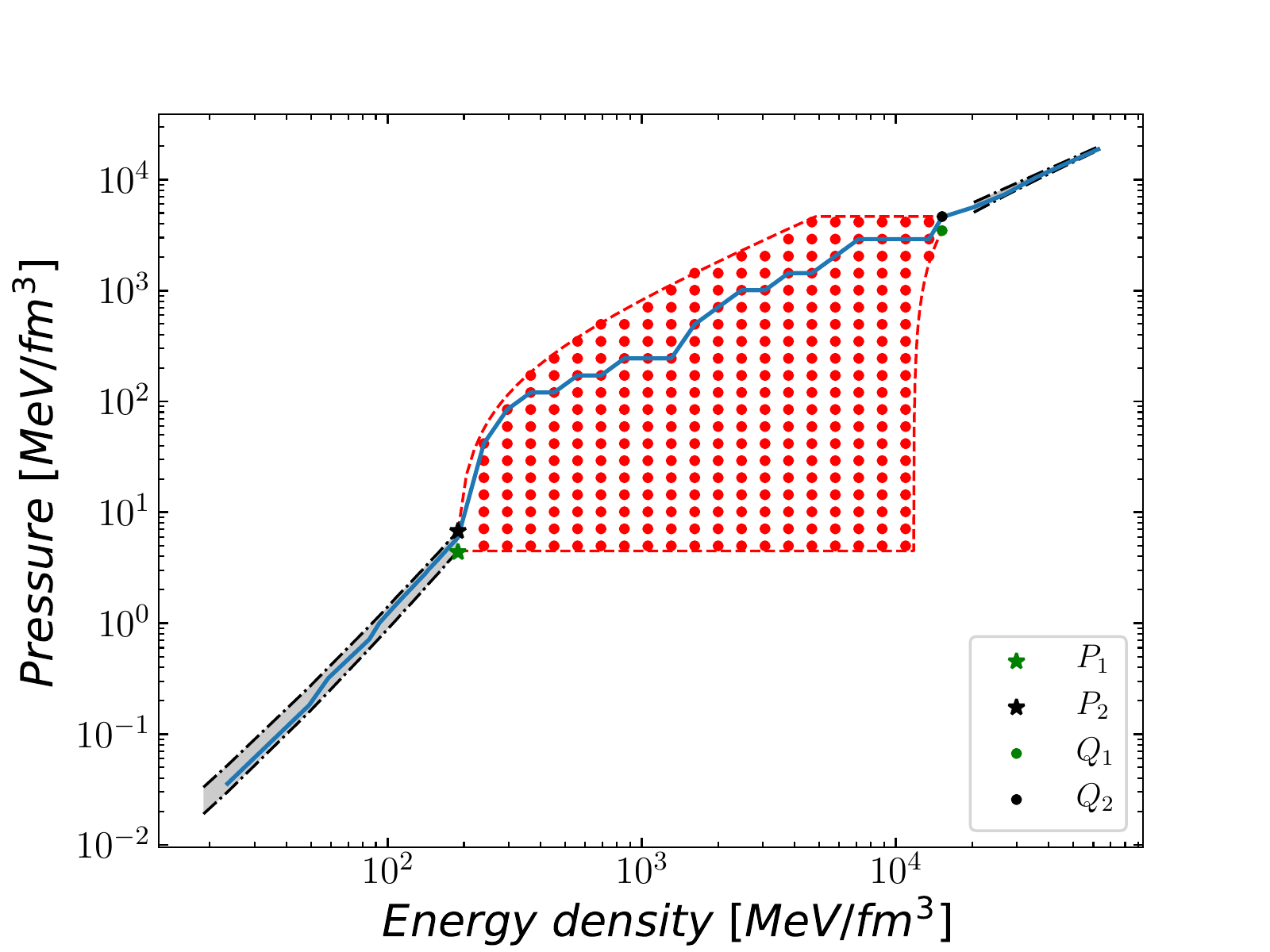}
\caption{{\bf Left:} simple sketch schematizing the construction of the boundary of the region at intermediate densities (as determined by causality and monotony) between chiral perturbation theory and high-density physics treated with pQCD (narrow bands at the left--bottom and right--top corner, respectively).
\label{fig:domain}
{\bf Right:} Additionally to the boundary  (red dashed lines) we show the nodes (red circles) of the grid interpolating at intermediate densities between the chiral computation of~\cite{Drischler:2016djf}  and the high--density physics treated with pQCD at $\mu_{\rm match} \sim 2.6$ (black dashed-dotted lines at the left--bottom and right top corner). A sample EoS is shown (solid line, blue online)).
%\label{fig:grid}
 }
\end{figure}

\begin{figure}[h]
\includegraphics[width=0.45\columnwidth]{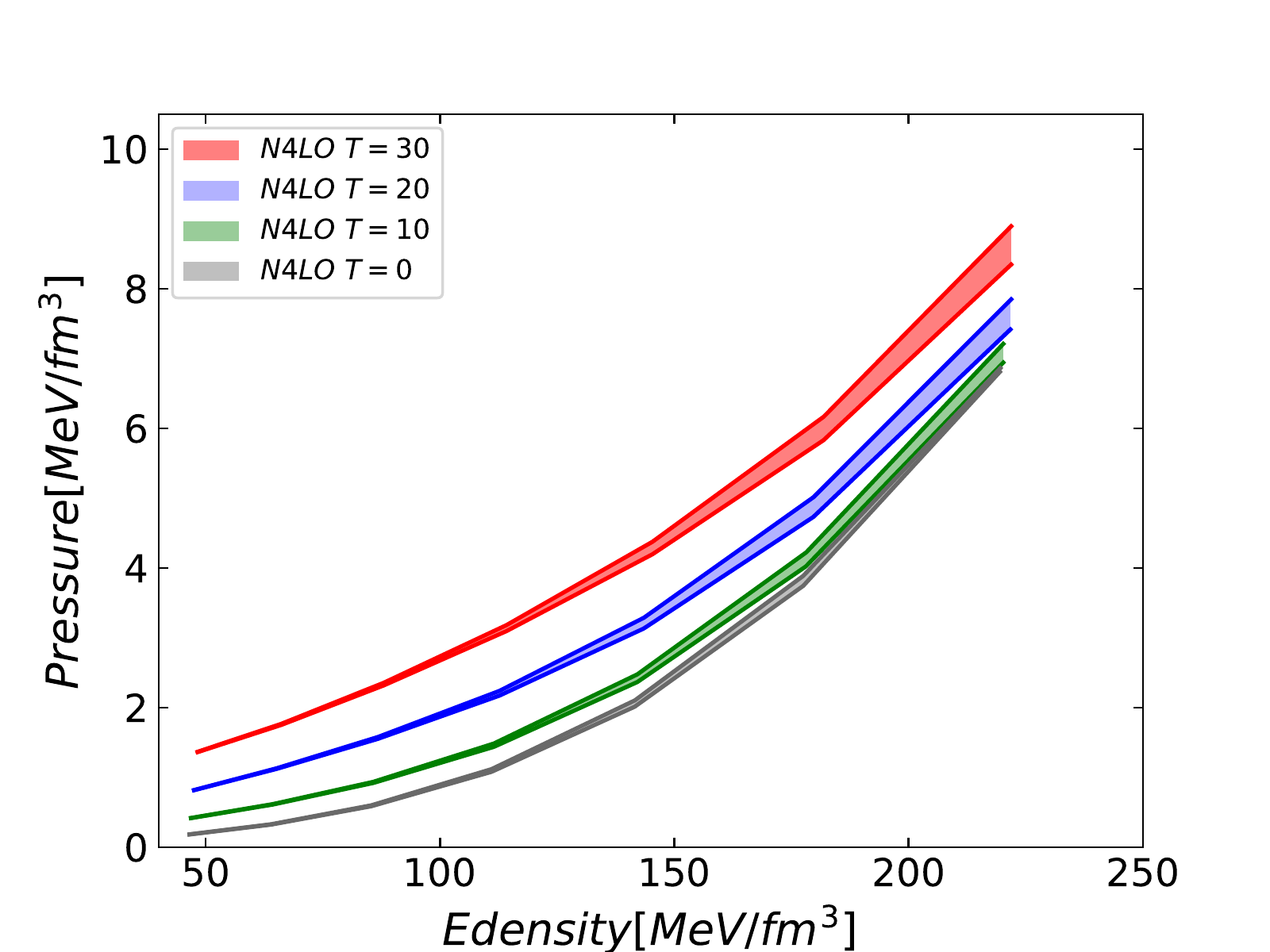}\ \ \ 
\includegraphics[width=0.45\columnwidth]{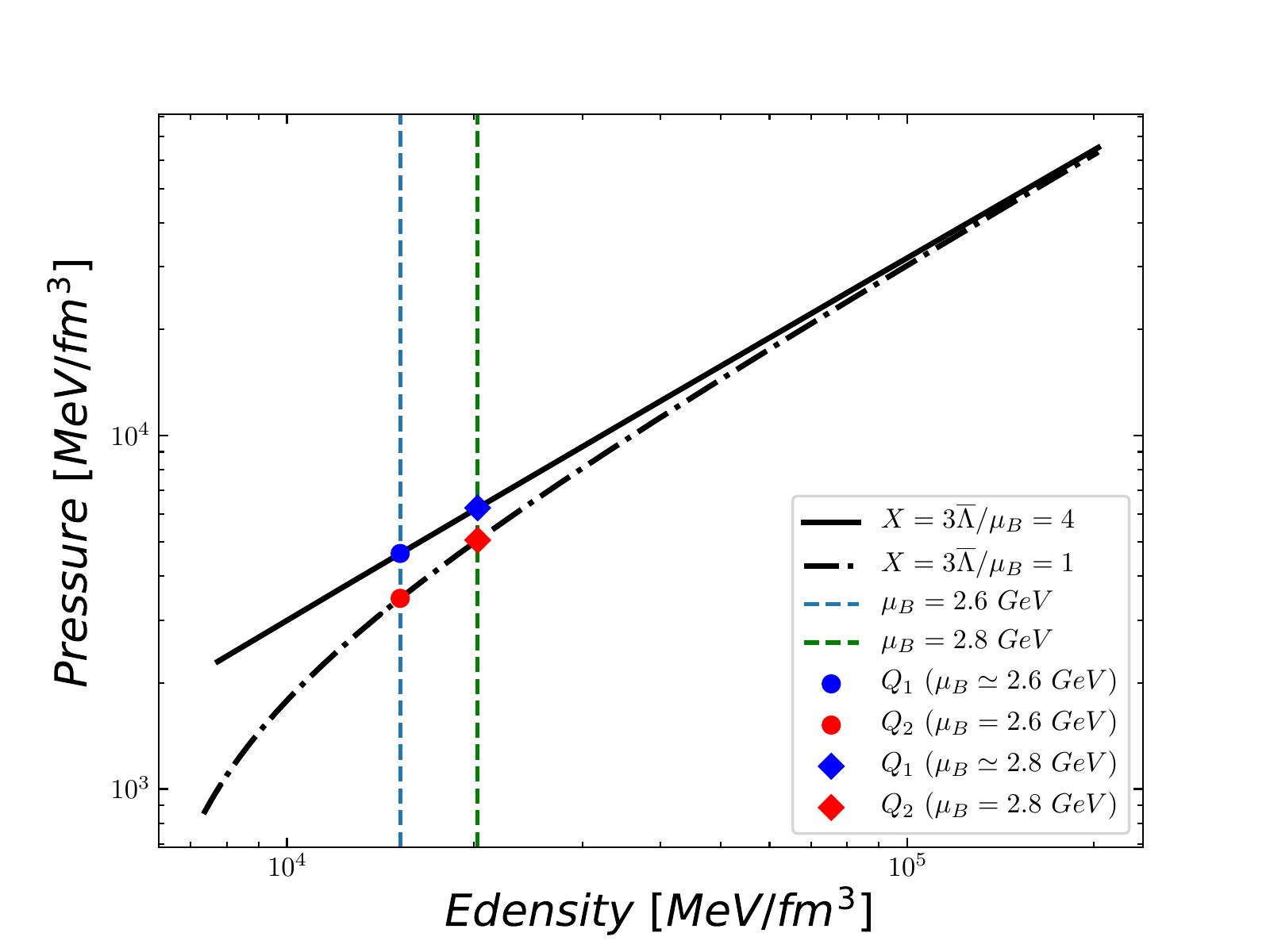}
\caption{Asymptotic limits for the EoS of neutron stars. 
{\bf Left:} neutron matter at low density, taken from the ChPT computations of the nuclear potential~\cite{Sammarruca:2020urk,Sammarruca:2021mhv,Sammarruca:2016ajl}, with $T$ increasing from bottom to top in 10 MeV steps. The effect of a typical temperature seems to be less than the uncertainty band.
{\bf Right:} High-density band from pQCD~\cite{Kurkela:2014vha} showing  $Q_1$ and $Q_2$ (for the two different sample values of $\mu_B$). 
They are picked in a high density range where the $P$ uncertainty is still narrow. Since we present $P(\varepsilon)$ and not $P(\mu_B)$, we first fix $Q_2$ from the higher ($X=4$, (in the notation of ~\cite{Kurkela:2014vha}) line with either $\mu_B=2.6$ or 2.8 GeV. We then use the resulting $\varepsilon$ to obtain a slightly modified $\mu_B$ and from the pQCD pressure we obtain the low $Q_1$ band end with $X=1$.
Lowering the chosen values of $Q_1$, $Q_2$ would better constrain the EoS at intermediate density at a formal level, but would make the perturbativity of QCD less-reliable. 
\label{fig:Kurkelagraph}
}
\end{figure}

 Basic properties of neutron stars such as the maximum allowed mass are most sensitive
 to the intermediate density range. Yet there,  QCD is not easily tractable: we need to resort to basic theoretical properties, causality ($c_s < c$) and monotony ($cs\geq 0$).
 Therefore, we have developed a set of EoS  reported in ~\cite{Oter:2019kig} to sample EoS at intermediate densities. We sample the uncertainty bands of Chiral Perturbation Theory (at low density), the band of perturbative QCD (at high density) and interpolate between both of them at intermediate density.
First, we fix  the low-density and high-density limits of the intermediate band as determined by monotony and causality, as shown in the left plot of  figure~\ref{fig:domain}. The low-density limits are determined by the different computations of ChPT ($P_1, P_2$ on that plot), while the choice of pQCD starting points (Q1 and Q2) is shown in the right plot of figure ~\ref{fig:Kurkelagraph}).

 Then, a grid is laid out in that band and for each energy density we choose a random value of the pressure from the values on the grid (right plot of figure~\ref{fig:domain}). Multiple transitions are possible as in complex condensed matter systems.
 
%%%%%%%%%%%%%%%%%%%%%%%%%%%%%%%%%%%%%%%%%%%%%%%%%%%%%%%% 
 \section{Latent heat of first-order phase transitions}
%%%%%%%%%%%%%%%%%%%%%%%%%%%%%%%%%%%%%%%%%%%%%%%%%%%%%%%%

The Gibbs thermodynamic-equilibrium condition determines at what critical chemical potential $\mu_c$  will the pressure of the two phases  be equal
\begin{equation}\label{Gibbscondition}
      T_H = T_E = 0 \ , \ \ \ \ 
      \mu  = \mu_c \ , \ \ \ \ 
      P_H = P_E := P_c\ .
\end{equation}

\begin{table}[h]
\centering
\caption{Characteristic examples of latent heat in known phase transitions (in natural units, $L$ being dimensionless). \label{tab:L} }
\begin{tabular}{|c|c|} \hline
Substance/transition  & $L$ \\ \hline 
He-3 superfluid                   & 1.5 $\mu$J/mol = $5.5\times 10^{-24}$\\ 
NdCu$_4$Fe$_4$O$_{12}$ perovskite & 25.5 kJ kg$^{-1}$ = $2.8\times 10^{-13}$\\
Ice-water             & 79.7 cal/g = $3.71\times 10^{-12}$ \\
Nuclear evaporation   & 30 MeV/A = $3\times 10^{-2}$ \\
Neutron star matter? & $>O(0.1)$? \\
\hline
\end{tabular}
\end{table}

First-order  phase transitions (PT) at zero temperature in neutron stars are characterized by a jump in the energy density  $\varepsilon_E - \varepsilon_H$ between the hadronic (H) and exotic (E) phases. This jump is usually quantified  by the specific latent heat of the transition L; table~\ref{tab:L}  shows some specific values for some cases, from condensed matter to nuclear physics~\cite{Carbone:2010ut}.

%%%%%%%%%%%%%%%%%%%%%%%%%%%%%%%%%%%%%%%%
\section{First order phase transition}
%%%%%%%%%%%%%%%%%%%%%%%%%%%%%%%%%%%%%%%%
The aim is to obtain L from the Equation of State (EoS) $P(\varepsilon)$ and our starting point is the latent heat per nucleon normalized to the vacuum neutron mass:
\begin{equation}\label{def:Ln}
L:=\frac{\Delta E}{NM_N}\ ,
\end{equation}

Integrating the first law of thermodynamics in terms of pressure we obtain the  energy per nucleon difference $\Delta E = E_E-E_H$ 

\begin{equation}\label{diff}
     \Delta E =P_H \frac{(n_E-n_H)}{n_E n_H}
\end{equation}
%allows to compute Eq.~(\ref{def:Ln}) for $L|_n$ that will be shown below in figure\ref{fig:L(n)}.O
Our approach is to use independent-model EoS by interpolating between low- and high-density regimes, but  the difficulty is not knowing  the energy per nucleon B/A and the density number in the area of intermediate energy from a matter Lagrangian. Therefore it has to be interpolated too, so as to quantify the latent heat.  Introducing in Eq.~(\ref{diff}) the energy density through
\begin{equation}\label{Eandn}
\varepsilon= n(M_N c^2+ B/A)\ ,
\end{equation}
the latent heat ($L|_\varepsilon$) can be obtained from the EoS 
\begin{equation}\label{LperA}
     L|_\varepsilon = P_H \frac{(\varepsilon_E-\varepsilon_H)}{\varepsilon_E  \varepsilon_H} \ .
\end{equation}

At low $\varepsilon$, Eq.~(\ref{LperA}) is equivalent to Eq.~(\ref{def:Ln}) save for the binding energy $0\simeq (B/A) << M_N$, introducing a few percent error quantified below. 
Near the density allowing maximum $L$, Eq.~(\ref{LperA}) is a very practical definition of $L$.

\begin{figure}[t]\vspace{-0.3cm}
\centering
\includegraphics[width=0.45\columnwidth]{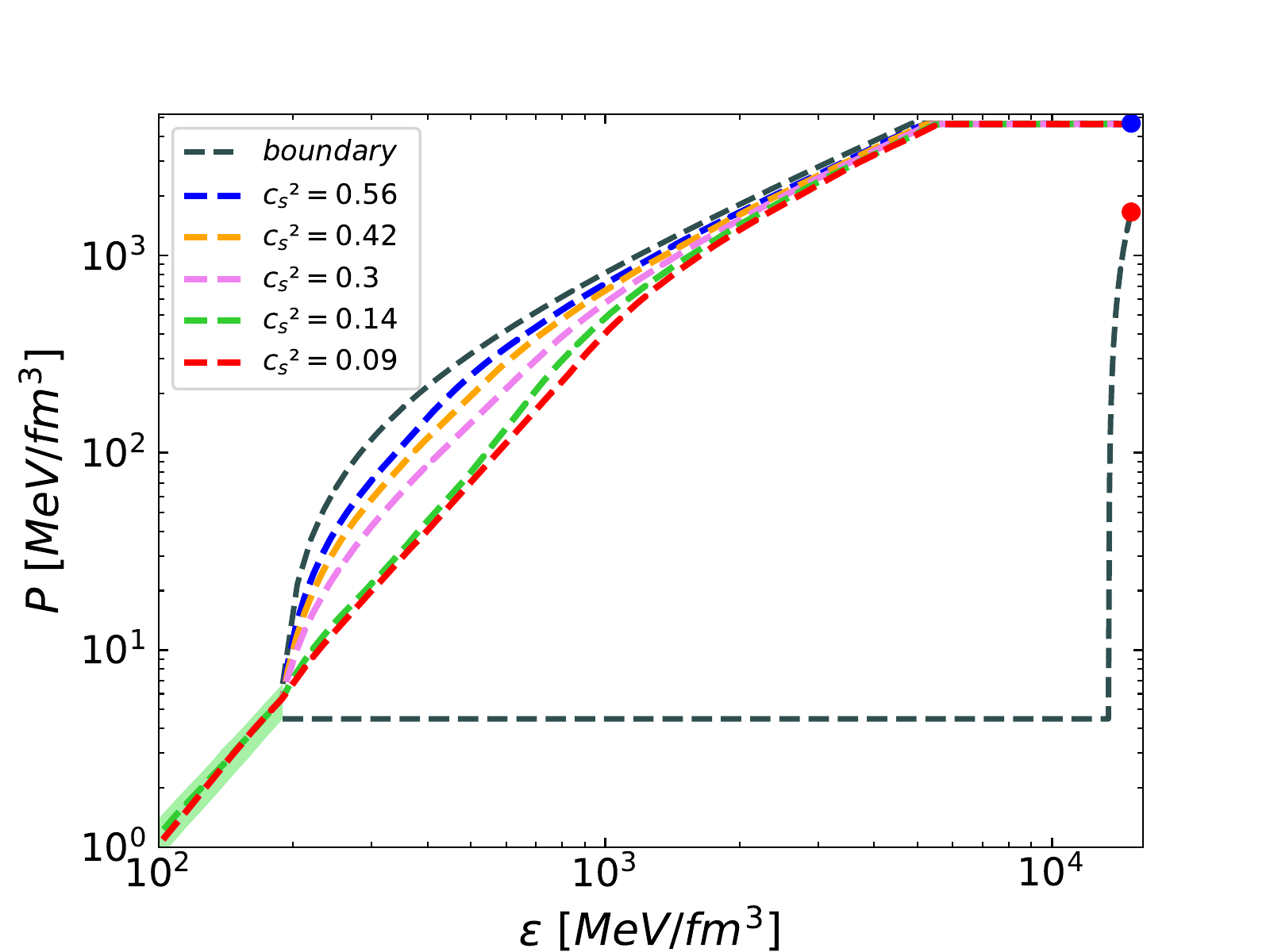}
\caption{\label{fig:band} nEoS~\cite{Oter:2019kig} theoretically allowed Eq. of state of neutron matter from hadron physics alone (pQCD and ChPT ~\cite{Drischler:2016djf,Kurkela:2009gj}) with no astrophysics nor General Relativity input, labelled in the same vertical order by $c_s^2$ at the ChPT matching point. The low-density ChPT approximation is used up to $1.2n_s$, just above the nuclear saturation density ($n_s\simeq 0.16 {\rm fm}^{-3}$).
\label{fig:EoS}
 }
\end{figure}

 Initially, we generated a set of EoS in order to look for the maximum latent heat in a PT, which are shown in Fig.~\ref{fig:band}.These are EoS of zero charge, $\beta$-stable neutron-star matter (NSM). 
At lowest number densities $n\leq 0.05 n_{\rm sat}$ (with saturation density $n_{\rm sat}\simeq 0.16/$fm$^3$) nuclear data directly  constrains the crustal EoS~\cite{Negele:1971vb,Baym:1971pw}. (Depending on the order of perturbation theory, this corresponds to $B/A\simeq 16\pm 1$ MeV added to the nucleon mass, and corresponds to a pressure around 153 MeV/fm$^3$).Shown here are the $N^3LO$ bands of~\cite{Drischler:2016djf,Drischler:2020yad} up to $n_m=1.2 n_{\rm sat}$.
The highest densities, at baryon chemical potential $\mu\geq 2.6\  GeV$ and above (this corresponds to an energy density of about 15 GeV/fm$^3$), can be studied with pQCD~\cite{Kurkela:2009gj}. It is likely that such high densities cannot be achieved in neutron stars within GR, but these limits  provide a powerful constraint on the shape to be adopted by the equation of state.
 
\label{sec-1}

The intermediate $\varepsilon$ region is enclosed in the figure by a soft-gray dotted line (occupying most of it in the log-scale employed). It is obtained by enforcing that the derivative of the EoS curves satisfy $0\leq P'(\varepsilon):= c^2_s \leq 1=c^2$ (that is, respecting monotony and causality). Thus, the speed of sound along the dotted boundary is either 0 or 1. 

We sample the contained region with a  500$^2$ grid.  Any $P(\varepsilon)$ in this region is matched to ChPT at a number density $n_m$, with random slope $P'=c^2_{s,m} \in (0,1)$ (given in the figure legends). At each successive grid point, another  random-slope step within those limits is taken.

For each one of this EoS, we computed the values for the specific latent heat $L$, starting by the simpler $L|_\varepsilon$. We proceeded by following each EoS that is compatible with all theoretical requirements (monotony and causality at every point, and satisfaction of both ChPT and pQCD constraints in their domains of validity), one at a time, from lower to higher $\varepsilon$.

At each grid point we computed the maximum stretch of $P'=0$ (first order phase transition) that could take place without violating any of those requirements, that is, we momentarily assume that very point to be the lower end of a phase transition, $(\varepsilon_H, P_H)$.

The largest maximum latent heat found for these conditions and with the current QCD understanding, 
$L|_\varepsilon\simeq 0.79$, would be reached for  $\varepsilon\simeq 11 \varepsilon_{\rm sat}$, (for an EoS matched to nuclear matter  with maximum slope $c^2_s\simeq 1$ at  1.2 $n_{\rm sat}$.

Assuming a small exotic matter core in the NS, this energy density discontinuity $\Delta \varepsilon$ has been quantified in terms of Seidov's limit ~\cite{Seidov:1971sv} 
\begin{equation}\label{Seidovsjump}
\Delta \varepsilon := \varepsilon_E-\varepsilon_H  = \varepsilon_H \left( \frac{1}{2} +\frac{3}{2}\frac{P_H}{\varepsilon_H} \right)
\end{equation}
which is based on how much energy density a neutron star core of the exotic phase transition can take before collapsing to a black hole (in the perturbative, small-core approximation). Thus, it is a property that depends on General Relativity. A comparison with our purely hadronic latent heat is adequate, but Seidov's should not be used with theories other than General Relativity.

\begin{figure}[h]
\centering
\includegraphics[width=0.45\columnwidth]{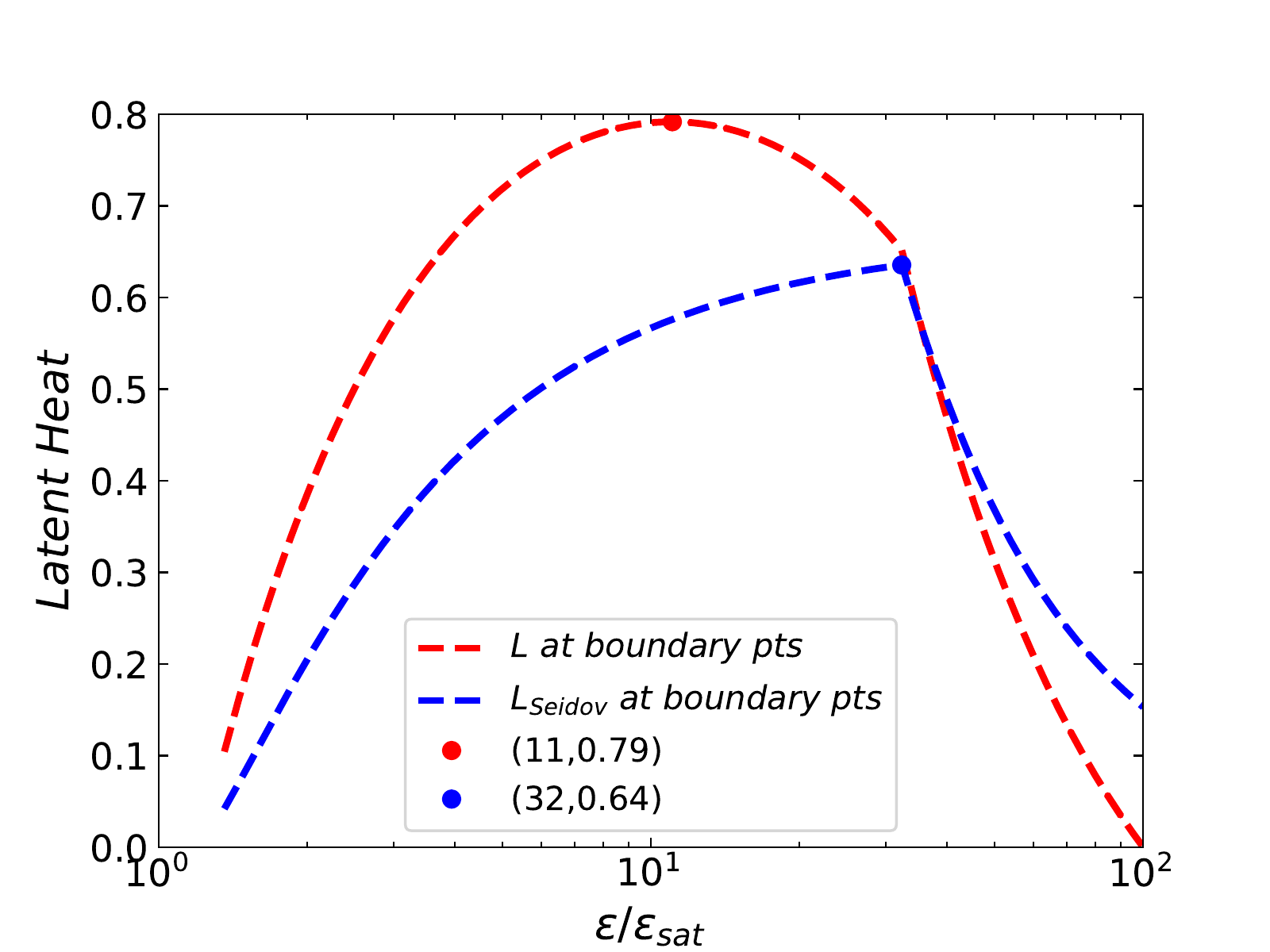}
\caption{We compare the bound on the latent heat (upper line, red online) from microscopic hadron physics alone with the maximum latent heat that a static body in equilibrium can tolerate in General Relativity, the Seidov limit~\cite{Seidov:1971sv}  (bottom line, blue online). 
\label{fig:compareSeidov}}
\end{figure}

To compare with the latent heat that we are discussing, we define an analogous quantity that we dub ``Seidov's Latent Heat'' ($L|_\varepsilon$) in the same way as in Eq~(\ref{diff}) taking  as $\Delta \varepsilon$ the value obtained from Seidov´s limit. Figure~\ref{fig:compareSeidov} shows results obtained for both latent heats $L$ and ($L_S$).

In our previous paper on this topic, the set of EoS was constructed to find the maximum latent heat in a PT, so that the stiffest equations possible have been generated by increasing smoothly slope up to $c_s^2=1$ and  then flattening, up to the maximum energy density allowed by the causality band limit in figure~\ref{fig:EoS}, as if a PT happened. 
Subsequently, a new work~\cite{Komoltsev:2021jzg} has shown that  new constraints must be taken into account, regarding the density number $n$ and the chemical potential $\mu$. These constraints derive from causality  in the $(n,\mu)$ plane instead of the $(\varepsilon,P)$ that contains the EoS, and impose a condition on the first derivative of the number density
\begin{equation}\label{constraint}
 c_s^{-2} \geq \frac{\mu}{n}\frac{\partial n }{\partial \mu}  \geq 1 \ .
\end{equation}

\begin{figure}
\includegraphics[width=0.6\columnwidth]{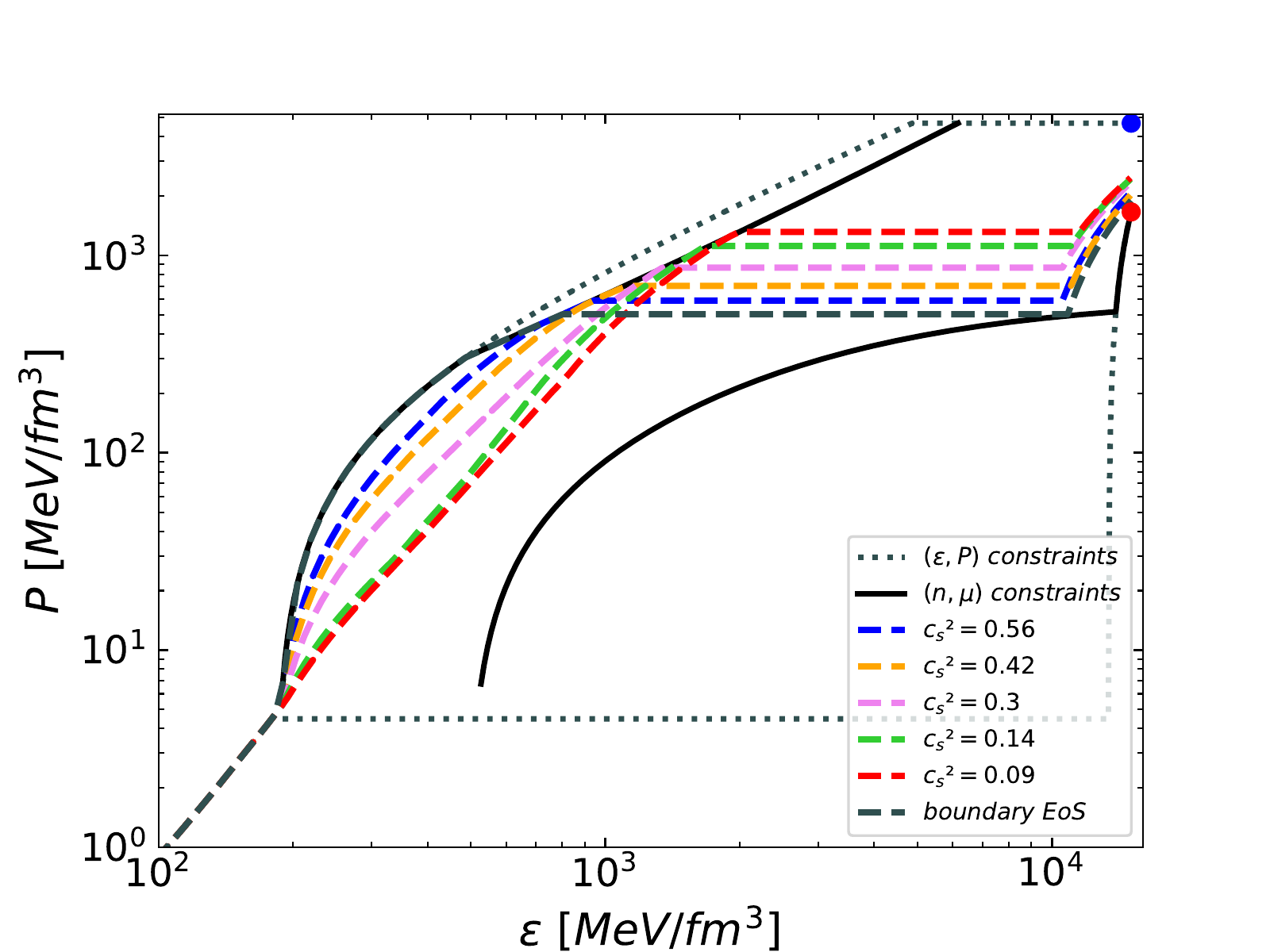}
\centering
\caption{nEoS~\cite{Oter:2019kig} theoretically allowed Eq. of state band of neutron matter from hadron physics alone (pQCD and ChPT ~\cite{Drischler:2016djf,Kurkela:2009gj}) with no astrophysics nor General Relativity input, labelled in the same vertical order by $c_s^2$ at the ChPT matching point, taking into account new constraints from ~\cite{Komoltsev:2021jzg}.
\label{fig:newboundary}}
\end{figure}

Consequently, we have to construct the allowed EoS by considering all possible functions $n(\mu)$, allowed by the above assumptions, connecting the low- and high-density limits related to the number density $n$ (or $\mu$). Taking into account these new constraints, the smaller boundary region obtained is shown in figure~\ref{fig:newboundary} and is an original contribution of this proceedings paper. Here, the new EoS  are constructed following the causality constraints  in both the $(\varepsilon, P)$ and $(\mu,n)$ planes simultaneously, as well as the limits of $P$ and $\mu$ at the starting pQCD regime. We have used the same criteria to construct all these set of EoS,  selecting for each density energy point of the grid the maximum pressure value allowed by the constraints to get the point where a PT happens, then flattening to the maximum energy density allowed. From this ending PT point to the starting pQCD regime, we construct the EoS applying the conform limit $c_s^2\simeq1/3$.

\begin{figure}
\centering
\includegraphics[width=0.45\columnwidth]{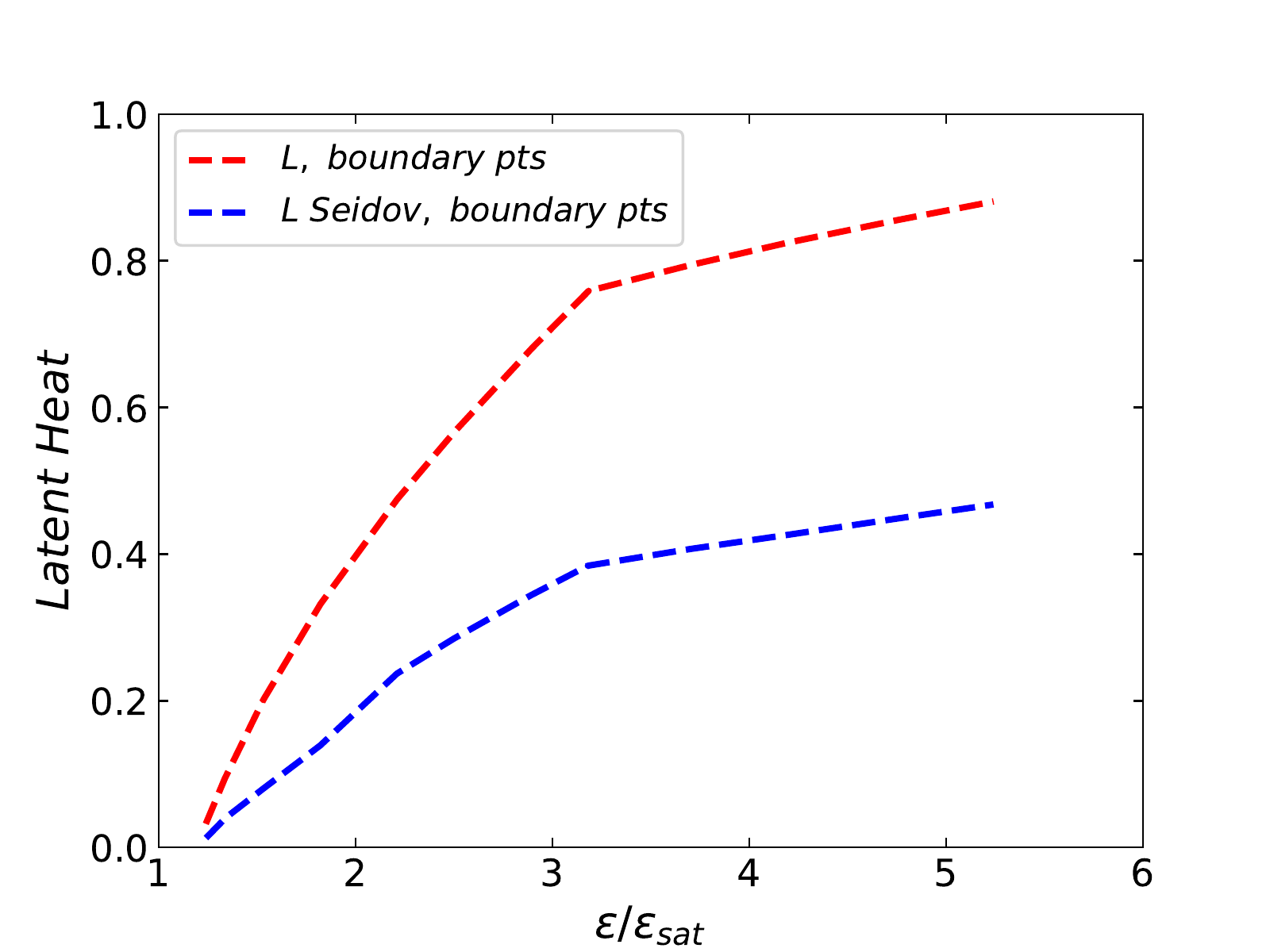}
\caption{We compare the bound on the latent heat (upper line, red online) from microscopic hadron physics alone (see Figure~\ref{fig:newboundary}) with the maximum latent heat that a static body in equilibrium can tolerate in General Relativity, the Seidov limit~\cite{Seidov:1971sv}  (bottom line, blue online) taking into account new constraints from the $(n,\mu)$ plane~\cite{Komoltsev:2021jzg}.  While Seidov's bound is better, it is a small-core approximation valid in General Relativity only; our less tight bound is of wider generality.
\label{fig:newcompareSeidov}
}
\end{figure}

The maximum $L$ resulting for the longest phase transition corresponding to the stiffest EoS in the entire construction (soft-grey dashed line) is $L= 0.88$. We compare  in Figure~\ref{fig:newcompareSeidov}  these results with the values obtained applying the Seidov's latent heat for the same EoS. There is  an appreciable difference between both values, indicating either that both pQCD and ChPT calculations have to be improved, or that this relativistic limit should be extended to not so small nuclei of neutron stars.

\begin{figure}
\includegraphics[width=0.45\columnwidth]{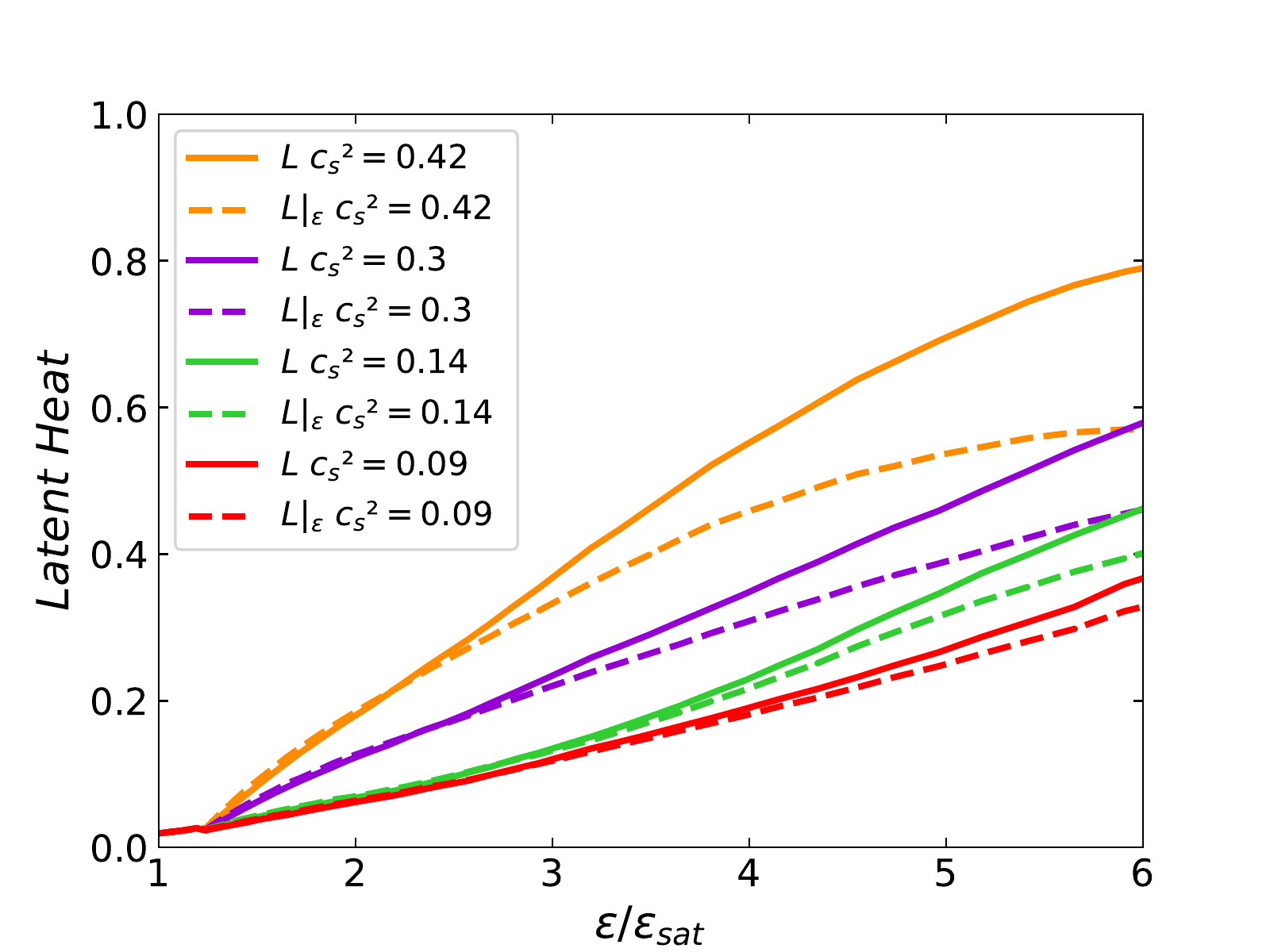}
\centering
\caption{We compare the two definitions $L$ and $L|_\varepsilon$ of Eq.~(\ref{def:Ln}) and Eq.~(\ref{diff}) respectively for the EoS with sound speeds ($0.42-0.09$) at the matching point. Below three times the energy density corresponding to nuclear saturation, one is in a nonrelativistic regime where the precise choice does not make a large difference. Above that, they start to diverge. If information from the $(n,\mu)$ plane is not at hand, that is, from $T^{\mu\nu}$ alone, our new definition in Eq.~(\ref{diff}) is more adequate.} 
\label{fig: LversusLE}
\end{figure}

Finally, in Figure~\ref{fig: LversusLE} we compare the two different definitions $L$ and $L|_\varepsilon$ in Eq.~(\ref{def:Ln}) and Eq.~(\ref{diff}) for some of these EoS. Around $\varepsilon=4.2\varepsilon_s$ our computations for the softest EoS differ between $4\%$  ($c_s^2=0.09$) and $12\%$ ($c_s^2=0.3$). For the stiffest EoS between ($c_s^2=1$ and $c_s^2=0.42$ ), the same variation of values happens at $\varepsilon=2.6\varepsilon_s$.   In the nonrelativistic limit they of course coincide.

%$

%L_nversusL_E_bdryConst

%%%%%%%%%%%%%%%%%%%%%%%%%%%%%%%%%%%%%%%%%%%%%%%%%%%%%%
\section{Conclusion}

Specific latent heat can be a useful quantity to characterize a phase transition in neutron stars. The results in this work show the sensitivity of this parameter to the construction of each EoS, which strongly depends on the choice of matching points at low and high density, due to the causality bounds at each point, as well as to the limitation of the maximum and minimum values ($\varepsilon, P,\mu, n$) of the entry points to the pQCD regime. The maximum latent heat obtained with this set of EoS is around 0.9 (in natural units). In a near future, as ChPT and pQCD calculations improve, these results will be better constrained, by quantifying the maximum possible phase transition allowed by  the uncertainty bands at low and high density. In fact, we have already updated our earlier results with some of the latest information  from Kurkela and Kolmotsev. 
Hadron physics predicts a maximum latent heat to any conceivable phase transition, even in theories that modify General Relativity.
%%%%%%%%%%%%%%%%%%%%%%%%%%%%%%%%%%%%%%%%%%%%%%%%%%%%%%

%%%%%%%%%%%%%%%%%%%%%%%%%%%%%%%%%%%%%%%%%%%%%%%%%%%%%%
\section*{Acknowledgment}
%%%%%%%%%%%%%%%%%%%%%%%%%%%%%%%%%%%%%%%%%%%%%%%%%%%%%%

I am thankful to Felipe J.Llanes-Estrada for assistance with the manuscript.
We have received support from spanish grants MICINN: PID2019-108655GB-I00, PID2019-106080GB-C21 (Spain); the COST action CA16214 (Multimessenger Physics and Astrophysics of Neutron Stars); Univ. Complutense de Madrid under research group 910309 and IPARCOS. 

\newpage

%%%%%%%%%%%%%%%%%%%%%%%%%%%%%%%%%%%%%%%%%%%%%%%%%%%%%%


\begin{thebibliography}{50}
%%%%%%%%%%%%%%%%%%%%%%%%%%%%%%%%%%%%%%%%%%%%%%%%%%%%%%

%\cite{Chesler:2019osn}
\bibitem{Chesler:2019osn}
P.~M.~Chesler, N.~Jokela, A.~Loeb and A.~Vuorinen,
%``Finite-temperature Equations of State for Neutron Star Mergers,''
Phys. Rev. D \textbf{100}, 066027 (2019)
doi:10.1103/PhysRevD.100.066027


\bibitem{Alford:2007xm}
M.~G.~Alford, A.~Schmitt, K.~Rajagopal and T.~Sch\"afer,
%``Color superconductivity in dense quark matter,''
Rev. Mod. Phys. \textbf{80}, 1455-1515 (2008)
doi:10.1103/RevModPhys.80.1455.


\bibitem{Fulde:1964zz}
P.~Fulde and R.~A.~Ferrell,
%``Superconductivity in a Strong Spin-Exchange Field,''
Phys. Rev. \textbf{135}, A550-A563 (1964)
doi:10.1103/PhysRev.135.A550;
F.~J.~Llanes-Estrada and G.~M.~Navarro,
%``Cubic neutrons,''
Mod. Phys. Lett. A \textbf{27}, 1250033 (2012)
doi:10.1142/S0217732312500332
%\cite{Hoffberg:1970vqj}
M.~Hoffberg, A.~E.~Glassgold, R.~W.~Richardson and M.~Ruderman,
%``Anisotropic Superfluidity in Neutron Star Matter,''
Phys. Rev. Lett. \textbf{24}, 775 (1970)
doi:10.1103/PhysRevLett.24.775


\bibitem{Oertel:2016xsn}
M.~Oertel, F.~Gulminelli, C.~Provid\^encia and A.~R.~Raduta,
%``Hyperons in neutron stars and supernova cores,''
Eur. Phys. J. A \textbf{52}, 50 (2016)
doi:10.1140/epja/i2016-16050-1

\bibitem{Heiselberg:1994fy}
H.~Heiselberg,
%``Quark matter structure in neutron stars,''
Contribution to the International Symposium on Strangeness and Quark Matter, 298-307
[arXiv:hep-ph/9501359 [hep-ph]].

%\cite{Llanes-Estrada:2019wmz}
\bibitem{Llanes-Estrada:2019wmz}
F.~J.~Llanes-Estrada and E.~Lope-Oter,
%``Hadron matter in neutron stars in view of gravitational wave observations,''
Prog. Part. Nucl. Phys. \textbf{109}, 103715 (2019)
doi:10.1016/j.ppnp.2019.103715



%\cite{Oter:2019kig}
\bibitem{Oter:2019kig}
E.~L.~Oter, A.~Windisch, F.~J.~Llanes-Estrada and M.~Alford,
%``nEoS: Neutron Star Equation of State from hadron physics alone,''
J. Phys. G \textbf{46}, 084001 (2019)
doi:10.1088/1361-6471/ab2567

%\cite{Godzieba:2020tjn}
\bibitem{Godzieba:2020tjn}
D.~A.~Godzieba, D.~Radice and S.~Bernuzzi,
%``On the maximum mass of neutron stars and GW190814,''
Astrophys. J. \textbf{908}, 122 (2021)
doi:10.3847/1538-4357/abd4dd


%\cite{Drischler:2016djf}
\bibitem{Drischler:2016djf}
C.~Drischler, A.~Carbone, K.~Hebeler and A.~Schwenk,
%``Neutron matter from chiral two- and three-nucleon calculations up to N$^3$LO,''
Phys. Rev. C \textbf{94}, 054307 (2016)
doi:10.1103/PhysRevC.94.054307



%\cite{Sammarruca:2020urk}
\bibitem{Sammarruca:2020urk}
F.~Sammarruca, R.~Machleidt and R.~Millerson,
%``Temperature effects on the neutron matter equation of state obtained from chiral effective field theory,''
Mod. Phys. Lett. A \textbf{35}, 2050156 (2020)
doi:10.1142/S0217732320501564

%\cite{Sammarruca:2021mhv}
\bibitem{Sammarruca:2021mhv}
F.~Sammarruca and R.~Millerson,
%``Analysis of the neutron matter equation of state and the symmetry energy up to fourth order of chiral effective field theory,''
Phys. Rev. C \textbf{104}, 034308 (2021)
doi:10.1103/PhysRevC.104.034308

%\cite{Sammarruca:2016ajl}
\bibitem{Sammarruca:2016ajl}
F.~Sammarruca, L.~Coraggio, J.~W.~Holt, N.~Itaco, R.~Machleidt and L.~Marcucci,
%``How well does the chiral expansion converge in nuclear and neutron matter?,''
PoS \textbf{CD15}, 026 (2016)
doi:10.22323/1.253.0026




%\cite{Kurkela:2014vha}
\bibitem{Kurkela:2014vha}
A.~Kurkela, E.~S.~Fraga, J.~Schaffner-Bielich and A.~Vuorinen,
%``Constraining neutron star matter with Quantum Chromodynamics,''
Astrophys. J. \textbf{789}, 127 (2014)
doi:10.1088/0004-637X/789/2/127



%\cite{Carbone:2010ut}
\bibitem{Carbone:2010ut}
A.~Carbone, A.~Polls, A.~Rios and I.~Vidana,
%``Latent heat of nuclear matter,''
Phys. Rev. C \textbf{83}, 024308 (2011)
doi:10.1103/PhysRevC.83.024308


%\cite{Kurkela:2009gj}
\bibitem{Kurkela:2009gj} 
  A.~Kurkela, P.~Romatschke and A.~Vuorinen,
  %``Cold Quark Matter,''
  Phys.\ Rev.\ D {\bf 81}, 105021 (2010)
  doi:10.1103/PhysRevD.81.105021.
  %%CITATION = doi:10.1103/PhysRevD.81.105021;%%



%\cite{Negele:1971vb}
\bibitem{Negele:1971vb}
J.~W.~Negele and D.~Vautherin,
%``Neutron star matter at subnuclear densities,''
Nucl. Phys. A \textbf{207}, 298-320 (1973)
doi:10.1016/0375-9474(73)90349-7


%\cite{Baym:1971pw}
\bibitem{Baym:1971pw}
G.~Baym, C.~Pethick and P.~Sutherland,
%``The Ground state of matter at high densities: Equation of state and stellar models,''
Astrophys. J. \textbf{170}, 299-317 (1971)
doi:10.1086/151216


%\cite{Drischler:2020yad}
\bibitem{Drischler:2020yad}  
C.~Drischler, J. A.~Melendez, R. J.~ Furnstahl and D. R.~Phillips,
%    title = "{Quantifying uncertainties and correlations in the nuclear-matter equation of state}",
   Phys. Rev. C {\bf 102} 054315,(2020)
    doi :10.1103/PhysRevC.102.054315
    

  
%\cite{Seidov:1971sv}
\bibitem{Seidov:1971sv}
Z. Seidov, Sov. Astron, 15 (347) (1971). 


%\cite{Komoltsev:2021jzg}
\bibitem{Komoltsev:2021jzg}
O.~Komoltsev and A.~Kurkela,
%``How perturbative QCD constrains the Equation of State at Neutron-Star densities,''
[arXiv:2111.05350 [nucl-th]].


%%%%%%%%%%%%%%%%%%%%%%%%%%%%%%%%%%%%%%%%%%%%%%%%%%%%%%%%%%%%%5


\end{thebibliography}
\end{document}